\documentclass{iaus}
\renewcommand{\vec}[1]{\mbox{\boldmath$#1$}}
\def\gsim{\lower.4ex\hbox{$\;\buildrel >\over{\scriptstyle\sim}\;$}}
\newcommand{\Om}{{\it \Omega}}
\usepackage{graphicx}

\title[Magnetic instability in  radiative  zones] 
{Magnetic pinch-type instability in stellar radiative  zones}

\author[G. R\"udiger, L. L. Kitchatinov \& M. Gellert]   
{G\"unther R\"udiger$^1$
 \break  Leonid L. Kitchatinov$^{1,2}$  \and Marcus Gellert$^1$}

\affiliation{$^1$Astrophysikalisches Institut Potsdam, An der Sternwarte 16, D-14482 Potsdam, Germany \break email: gruediger@aip.de\\[\affilskip]
$^2$Institute  for Solar-Terrestrial Physics, P.O. Box 291, Irkutsk, 664033, Russia}

\pubyear{2008}
\volume{259}  
\pagerange{1--}
\date{?? and in revised form ??}
\jname{Cosmic Magnetic Fields:
from Planets, to Stars and Galaxies}
\editors{ K.G. Strassmeier,
A.G. Kosovichev \&
J. Beckmann, eds.}
\begin{document}

\maketitle

\begin{abstract}
The solar tachocline is shown as  hydrodynamically stable against nonaxisymmetric disturbances if it is true that no cos$^4\theta$ term exists 
in its rotation law. We also show that the toroidal  field of 200 Gauss amplitude which produces the tachocline in the 
magnetic theory of R\"udiger \& Kitchatinov (1997) is  stable  against nonaxisymmetric MHD disturbances -- but 
it becomes unstable  for rotation periods slightly slower than 25 days. The instability of such  weak fields  
lives from the high thermal diffusivity   of  stellar radiation zones compared with the magnetic diffusivity. The  
growth times, however, result as  very long (of order of 10$^5$ rotation times). With estimations of the chemical mixing  we find  the maximal possible field  amplitude  to be  $\sim$500 Gauss in order to explain   the observed lithium abundance of the Sun. Dynamos with such low field amplitudes should not be relevant for the solar activity cycle.

With nonlinear simulations of  MHD Taylor-Couette flows  it is shown that for the rotation-dominated magnetic instability  
the resulting eddy viscosity  is only of the order of the molecular viscosity. The Schmidt number as the ratio of viscosity and chemical diffusion   grows to values of $\sim 20$. For the majority of the stellar physics  applications, the magnetic-dominated Tayler instability will be quenched by the stellar  rotation.

\end{abstract}

\firstsection 
\section{Introduction}
We ask for the stability of differential rotation in radiative stellar zones  under the presence of magnetic fields. If the magnetic field is aligned with the rotation axis then the answer is simply `magnetorotational instability' (MRI). If the field is mainly toroidal (the rule rather than the exception) the answer is more complicated. Then the Rayleigh criterion for stability against axisymmetric perturbations of Taylor-Couette (TC)  flow with the rotation profile $\Om=\Om(R)$ reads
\begin{equation}
\frac{1}{R^3}\frac{\rm d}{{\rm d} R}(R^2 \Om)^2 - \frac{R}{\mu_0\rho}\frac{\rm d}{{\rm d}R} \left(\frac{B_\phi}{R}\right)^2>0.
\label{RC}
\end{equation}
Hence, almost uniform fields or fields with $B_\phi\propto 1/R$ (a current-free field) are {\em stabilizing} the TC flow and no new instability appears.

More interesting is the question after the stability  against nonaxisymmetric perturbations. Tayler (1973) found the necessary and sufficient condition  
\begin{equation}
\frac{\rm d}{{\rm d} R}(R B^2_\phi)<0
\label{TC}
\end{equation}
for stability of an ideal fluid against nonaxisymmetric perturbations.  Now almost homogenous fields are unstable while the fields with $B_\phi\propto 1/R$ are stable.  

We have probed the interaction of such stable toroidal fields with stable flat rotation laws and found, surprisingly, the Azimuthal Magnetorotational Instability (AMRI)  which for small magnetic Prandtl number  scales with the magnetic Reynolds number Rm of the global rotation similar to the standard MRI (R\"udiger et al. 2007). 

In the following, as an astrophysical application of these nonaxisymmetric instabilities the magnetic theory of R\"udiger \& Kitchatinov (1997, 2007)  of the solar tachocline is presented. In the last Section we return to a TC  flow under the presence of strong enough toroidal fields presenting first results of the eddy viscosity and the turbulent diffusion of chemicals for the Tayler instability (TI).
\section{Solar tachocline}
The tachocline is the thin shell between the solar convection zone and the radiative interior of the Sun where the rotation pattern dramatically changes.
\begin{figure}[h]
\begin{center}
 \includegraphics[height=6cm, width=13cm]{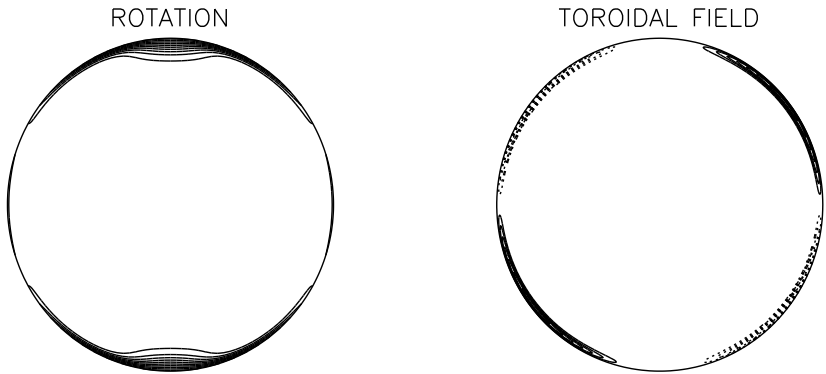}
  \caption{The tachocline formation on the basis of a fossil poloidal field of 10$^{-4}$ Gauss confined in the radiative solar interior  (R\"udiger \& Kitchatinov 1997). Left: the stationary rotation profile; right: the isolines of the resulting toroidal field with its amplitude of ~200 Gauss.}\label{f0}
\end{center}
\end{figure}

The nonuniform rotation of the solar convection zone -- which is due to the interaction of the convection with the global rotation -- has no counterpart in the solar core but the convection zone rotates in the average  with the same  angular velocity as the interior does. The radial coupling is thus large. This phenomenon cannot be explained by viscous  coupling (the viscosity  below the convection zone is by  more than 10 orders of magnitude smaller) but it can be explained with a weak fossil  poloidal field which is confined in the solar radiative interior. For the amplitude of this field  only values  of order mGauss are necessary  resulting in a  tachocline thickness of about 5\% of the solar radius (Fig.~\ref{f0}). The resulting toroidal field amplitude inside the tachocline of about 200 Gauss mainly depends on the magnetic Prandtl number (and the rotation velocity) for which  ${\rm Pm}=5\cdot 10^{-3}$ has been used in the model (R\"udiger \& Kitchatinov 1997, 2007). One can estimate the resulting toroidal field in terms of the Alfv\'en velocity $V_{\rm A}=B_\phi/\sqrt{\mu_0\rho}$ simply as
\begin{equation}
V_{\rm A}=\sqrt{\rm Pm} U_0
\label{VA}
\end{equation}
with $U_0$ the linear velocity of rotation, $U_0=R\Om$. For ${\rm Pm}\simeq 1$ the resulting toroidal field amplitude would be  of order 100 kGauss which is certainly unstable. For ${\rm Pm}\simeq 10^{-4}$ the field strength is reduced to only 1 kGauss so that  we have carefully to check its stability.
The magnetic theory  holds for two main conditions: i) the field must completely be confined in the radiation zone 
and ii)  the magnetic Prandtl number must be small enough.
There are several possibilities to fulfill the first condition (cf. Garaud 2007; R\"udiger \& Kitchatinov 2007) which, however,  shall not be discussed in the present paper.
\subsection{Hydrodynamic stability}
To fulfill the second condition the radiative tachocline must {\em hydrodynamically} be stable. On the first view this should not be a problem. As the differential rotation in latitude forms a shear flow its  amplitude,  $\delta\Om=\Om_{\rm eq}-\Om_{\rm pole}$, decides the stability properties. By use of a 2D approximation which ignores the radial coordinate Watson (1981) derived for ideal fluids the condition $\delta\Om/\Om < 0.286$ for stability. This  rather large value   would lead to a stable  tachocline. 
\begin{figure}(h)
\begin{center}
 \includegraphics[height=4.5cm,width=8cm]{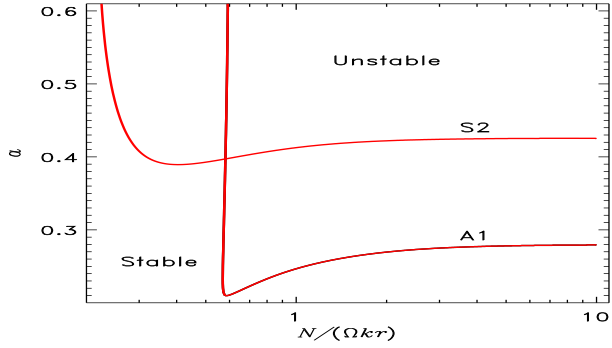}
  \caption{Neutral hydrodynamic instability lines for the rotation law without cos$^4\theta$-term in 3D. Only A1 and S2 modes are unstable. The A1 mode for  wavenumber $k\to 0$ reproduces the Watson (1981) result $a=\delta\Om/\Om_0\simeq 0.286$.}\label{f1}
\end{center}
\end{figure}
The radial velocity  components, however,  are small but not zero. Also the latitudinal profile of the angular velocity is more complicate than the simple $\cos^2\theta$-law used by Watson. We have thus to   rediscuss the stability of shear flows with 
\begin{equation}
\Om(\theta)=\Om_0\bigg(1-a((1-f)\cos^2\theta + f \cos^4\theta)\bigg)
\label{Omtheta}
\end{equation}
where $\delta\Om/\Om_0=a$; $f$ is the contribution of the $\cos^4\theta$ term which describes the shape  of the rotation law in midlatitudes. At the solar surface we have $a\simeq 0.286$ and $f=0.55$. 

The radial density gradient forms  a `negative' buoyancy leading to damped  oscillations with the frequency
\begin{equation}
N=\left(\frac{g}{C_p}\frac{\partial S}{\partial r}\right)^{\frac{1}{2}}.
\label{N}
\end{equation}
 $S$ is the entropy. The frequency $N$ in the upper  solar core is by more than a factor of 100 larger than the rotation frequency $\Om$.
So the equation system
\begin{eqnarray}
&& \frac{\partial \vec{u}}{\partial t}+(\vec{U}\nabla) \vec{u}+ (\vec{u} \nabla)\vec{U}= -\left(\frac{1}{\rho} \nabla {p}\right)'+ \nu\Delta \vec{u}\nonumber\\
&& T\left(\frac{\partial s}{\partial t}+(\vec{U}\nabla)s + (\vec{u}\nabla) S\right)= C_p\chi \Delta T'
\label{eqsyst}
\end{eqnarray}
must be solved for the flow perturbation $\vec{u}$ and  the entropy fluctuation $s$.  It is  ${\rm div} \vec{u}=0$; $s=-C_p \rho'/\rho$. $T'$  and $\rho'$ are the fluctuations of  the temperature and the  density, resp. The mean flow  $\vec{U}$ is given by (\ref{Omtheta}). The equations are solved with a Fourier expansion ${\rm exp}({\rm i}(kr+m\phi - \omega t))$ in the short-wave approximation $kr\gg 1$ with   $m$  as the azimuthal wave number. As usual, in   latitude a series expansion after  Legendre polynomials is used.

In both latitude and longitude the modes are  global. The parameter  including  the density stratification is
\begin{equation}
\hat\lambda= \frac{N}{kr\Om},
\label{hatlam}
\end{equation}
so that   $\hat\lambda\to \infty$ reproduces the 2D approximation by Watson (1981)  and Cally (2001). They showed that only nonaxisymmetric modes with $m=1$ can be unstable and the same is true in the present 3D approximation. 
The modes with $u_\theta$ antisymmetric with respect to the equator are marked with  Am and the modes with $u_\theta$ symmetric with respect to the equator are marked with  Sm . 

Let us start with the  rotation law ($f=0$). The Prandtl number is fixed as 
\begin{equation}
{\rm Pr}= \frac{\nu}{\chi}=2\cdot 10^{-6},
\label{Pr}
\end{equation}
where $\nu$ and $\chi$ are viscosity and thermal conductivity.

The main result is given in Fig.~\ref{f1} which shows the neutral-stability lines for various $a=\delta\Om/\Om$. Only A1 and S2 are obtained as unstable (S1 is stable!). For $k\to 0$ the  Watson result $a=0.286$ is reproduced. With radial stratification, however, this critical value is reduced to $a=0.21$. For ideal fluids Cally (2003) found instability for $a=0.24$ which also fits  our result. Hence, for $a<0.21$ the solar tachocline remains  hydrodynamically stable. There is thus no shear-induced turbulence. Note also that for $N\to 0$ (mimicking the convection zone)  no instability exists.
\begin{figure}[htb]
\begin{center}
\includegraphics[width=10cm,height=6.5cm]{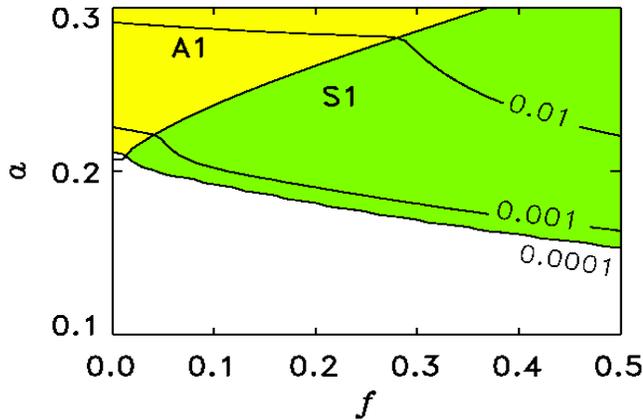}
  \caption{Hydrodynamic-stability map for various values $f$ for  the power of the $\cos^4\theta$-term in the rotation law (\ref{Omtheta}). This term  destabilizes  the S1 mode. For high $f$ already rather small shear  values become unstable.}\label{f2}
\end{center}
\end{figure}

The calculations have been repeated with the $\cos^4\theta$-term included in the rotation law. Then the S1 mode becomes dominant and   reduces the critical shear. For $f=0.5$ the maximum shear for stability results to  $a=0.16$ (Fig.~\ref{f2}). 

This is a rather small value. Both our modifications of the Watson approach are destabilizing the differential rotation. If the shape of the surface rotation with $f=0.5$ would be  conserved through the convection zone {\em and} the tachocline  then only 50\% of the surface shear  can indeed produce hydrodynamic turbulence in the tachocline. If the shear is not conserved ($f=0$) then the minimum shear for turbulence is 0.21,  large enough to ensure stability.

A dramatic {\em stabilization}, however, of the latitudinal shear results from the inclusion of the tachocline rotation law as observed  with its large radial gradients into the calculations.  With a 3D code without stratification one can  show that in this case all rotation laws with $a<0.5$ should be stable (Arlt et al. 2005).

In order to decide the stability problem of the solar tachocline informations are needed about the exact shape of the rotation law  there. Charbonneau et al. (1999) analyzed the helioseismic data and found $a\leq  0.15$ and $f\simeq 0$. After our analysis (and also after theirs) such a rotation law is hydrodynamically {\em stable}. Obviously, we  have to know in detail   the space dependence (and time dependence) of the internal solar rotation.

\subsection{MHD stability}
We consider  the tachocline  as hydrodynamically stable. The magnetic Prandtl number  is thus the microscopic  one for which we use ${\rm Pm}=0.005$ as a characteristic value. For the instability of toroidal fields the ratio of Pm and Pr, the Roberts number 
\begin{equation}
{\rm q}= \frac{\chi}{\eta}
\label{q}
\end{equation}
plays a basic role. For the Sun the typical  value   2500 is used. The interaction of   rotation  and toroidal magnetic  fields (of a simplified structure)  will be  demonstrated   with these parameters. The general result is  that  for small $\rm q$ the field must be strong to become unstable  while for $\rm q\gg 1$   much weaker fields become unstable. However, the growth times (in units of the rotation period)  for the weak-fields are much longer  (by  orders of magnitudes) than the growth times  of strong fields (of order of the   Alfv\'en period).

\begin{figure}
\hskip-0.5cm
\hbox{
\includegraphics[width=7cm,height=5cm]{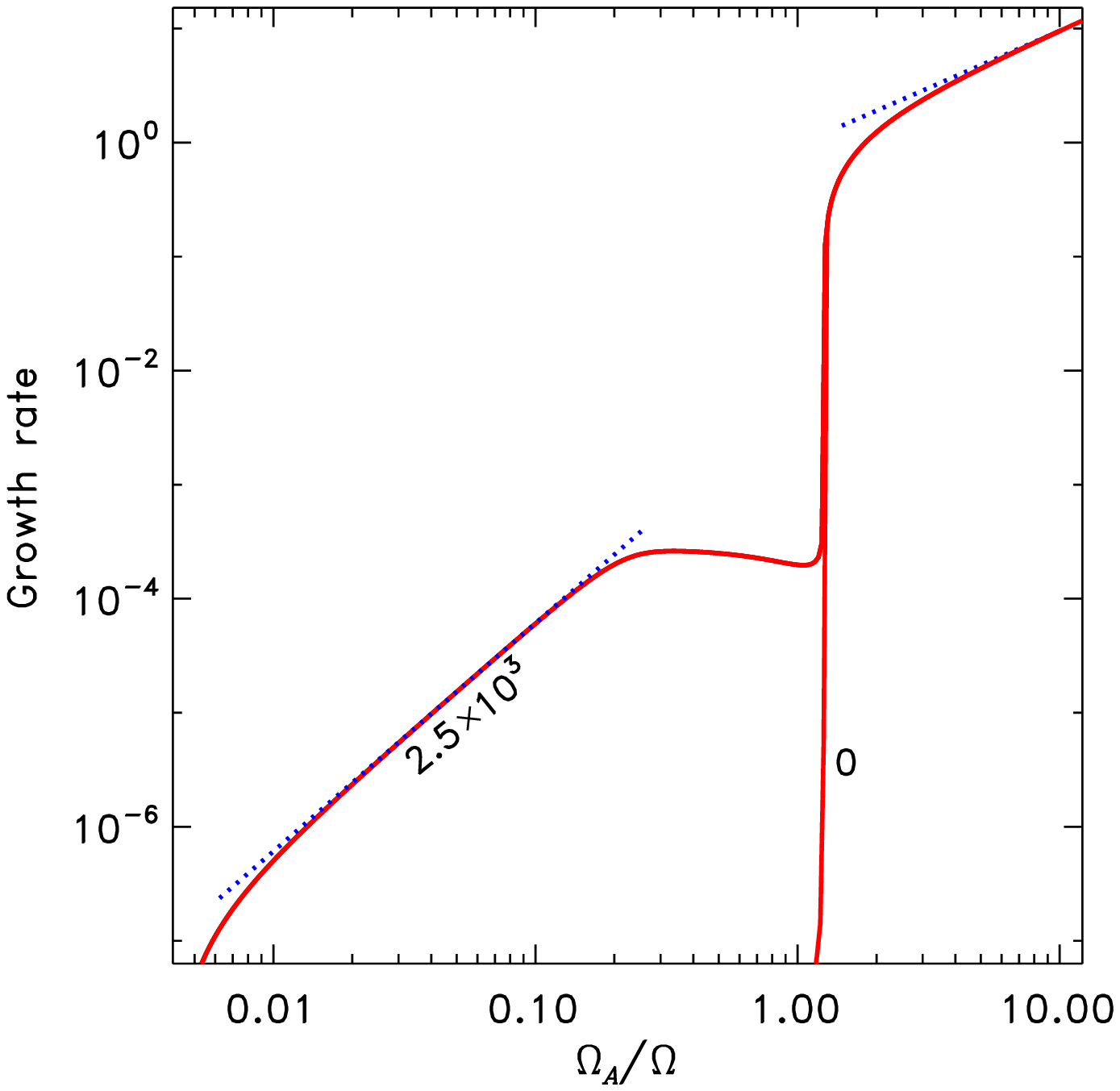}
\includegraphics[width=7cm,height=5cm]{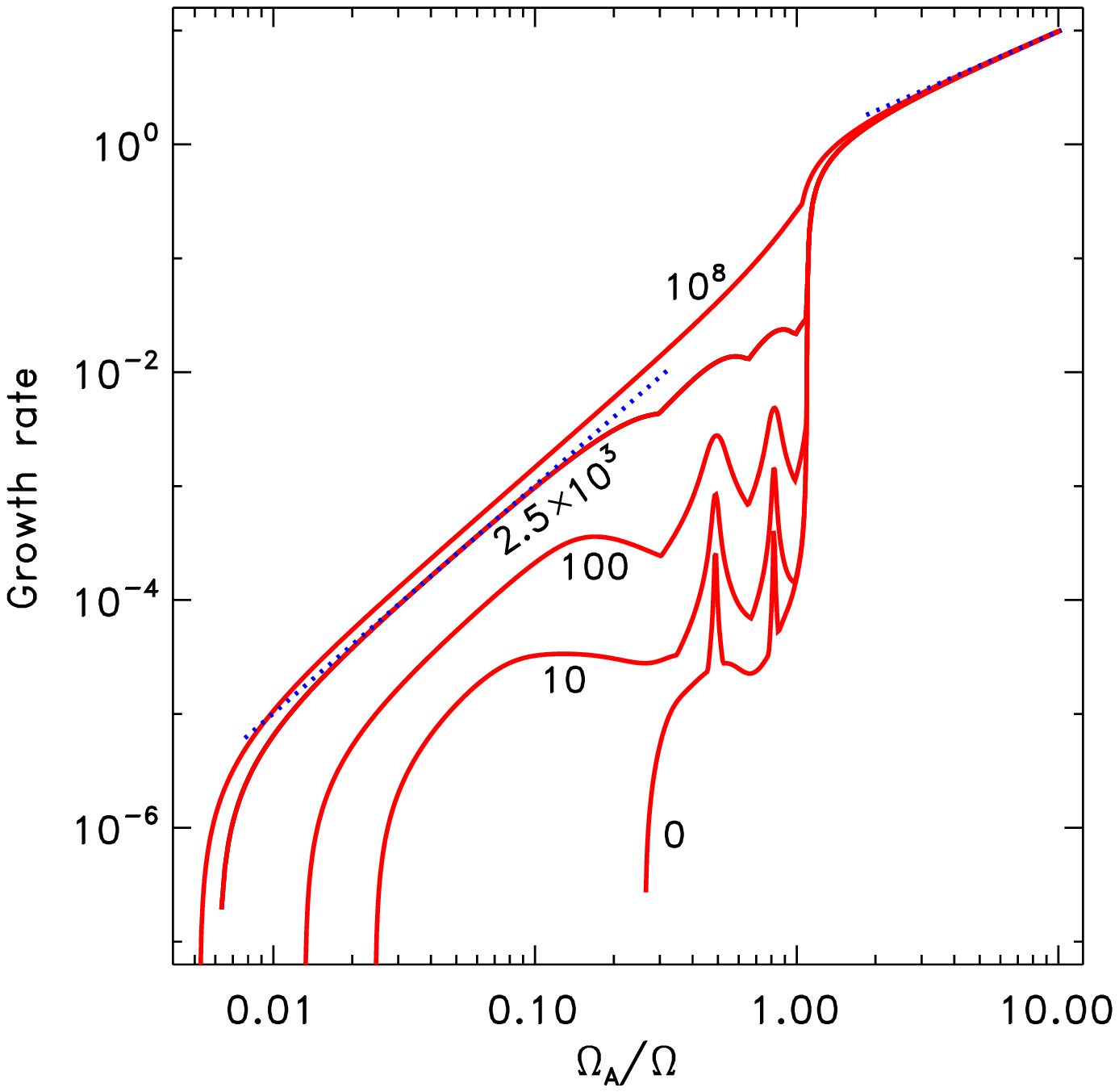}}
  \caption{Normalized growth rates for the magnetic instability in rigidly rotating stars. The curves are marked with their values of the Roberts number (\ref{q}). Left: one magnetic belt peaking at the equator; right: two magnetic belts with zero-field at the equator. There are  two instability domains: for strong magnetic fields ($\Om_{\rm A}>\Om$) the growth rates are high ($\simeq \Om_{\rm A}/\Om$) and for weak magnetic fields ($\Om_{\rm A}<\Om$) they are very small ($\simeq (\Om_{\rm A}/\Om)^2$). The latter domain only exists for $\rm q \gg1$ (perfect heat conduction).}\label{f3}
\end{figure}

 The left plot of  Figs.~\ref{f3} concerns a field with only one belt which peaks at the equator. The Alfv\'en frequency $\Om_{\rm A}$ is defined by
\begin{equation}
B_\phi= r \sin\theta \sqrt{\mu_0\rho} \Om_{\rm A},
\label{Bphi}
\end{equation}
and  is considered as constant (see Cally 2003). The global rotation is assumed as rigid. For this case Fig.~\ref{f3} (left) reveals the fields with $\Om_{\rm A}>\Om$ as always unstable with growth rates $\gsim\!\Om_{\rm A}$. Toroidal fields  with amplitudes $V_{\rm A} \gsim U_{\rm eq}$, i.e.  $\sim$10$^5$ Gauss (for the Sun) cannot stably exist in the radiative solar core. Even weaker fields with $\Om_{\rm A}\gsim 0.005 \Om$ can  be unstable but only for very high  heat-conductivity. For $\rm q=0$ the weak fields   are stable. For $\rm q\gg 1$ they are indeed unstable but with growth rates smaller than $10^{-4}$. The growth time $\tau_{\rm growth}$ in  units of the rotation period is
\begin{equation}
\frac{\tau_{\rm growth}}{\tau_{\rm rot}}= \frac{1}{2\pi\gamma}
\label{tautau}
\end{equation}
with the normalized growth rate  $\gamma=\Im(\omega)/\Om_0$. The maximum growth time is then  $>1000$  rotation periods (but much shorter than the diffusion time).  Even for very large $\rm q$ there is a magnetic limit below   the magnetic field is always stable. From Fig.~\ref{f3} (one or two belts) we find  the minimum field is $\Om_{\rm A}\simeq 0.005 \Om$. For the Sun, therefore, the maximum  stable field in the model is $\sim$600 Gauss.

The model is insofar  correct as the microscopic diffusivity values (viscosity, magnetic diffusivity, heat-conductivity)   have their real amplitudes  (for  ideal fluids, see Cally 2003). The model, however,   is insofar not correct  as the radial  profiles of the fields  are assumed as nearly  uniform. Arlt et al. (2007) work  with a 3D code without buoyancy ($\rm q \to \infty, Pm =0.01$)  for toroidal field belts with strong radial gradients and  find instability for weak fields with amplitudes of order 10 Gauss.

Not surprisingly, for  two  belts with equatorial antisymmetry (the field vanishes at the equator)  there are some differences to the one-belt model, but  the maximal stable field  amplitudes  are always of the same order (Fig.~\ref{f3}, right). 

%

\subsection{Effect of stellar spin-down}
\begin{figure}
\begin{center} 
\includegraphics[width=10cm,height=6cm]{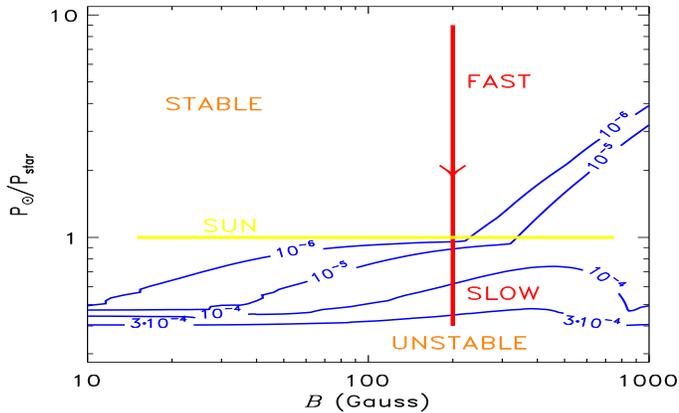}
  \caption{A star  spinning down moves from the top (stable area) to the bottom (unstable area). It is shown that a toroidal magnetic field of 200 Gauss is stable for the solar rotation period but the stability is lost for slower rotation.}\label{f4}
\end{center}
\end{figure}

The question arises whether  a solar-type star is always able  to form a tachocline. The older the star the slower  its rotation. Hence the rotational quenching of the Tayler instability becomes weaker and weaker  for older stars so that the instability becomes more efficient. By its spin-down the star moves to the right along the abscissa of both the Figs.~\ref{f3}. The (slow) magnetic decay goes in the opposite direction; this effect is still neglected. We assume that the total amount of the latitudinal differential rotation remains constant during the star's spin-down. This is a well-established assumption (see Kitchatinov \& R\"udiger 1999; K\"uker \& Stix 2001).

Figure~\ref{f4} shows the results. We have computed the normalized growth rates of the magnetic instability of rotating stars with a toroidal field of 200 Gauss and a differential rotation of $\delta \Om\simeq 0.06$ day$^{-1}$ (the solar value). The rotation period is normalized  with 25 days in Fig. \ref{f4} so that the horizontal   yellow line represents the Sun. In the upper part of the plot the 200 Gauss are stable while in the lower part of the plot they are unstable. Obviously, the Sun lies in the stable area but very close to the instability limit. We are thus tempted to predict that G2 stars older than the Sun (or better: of slower rotation) should not have a tachocline. When the toroidal field becomes unstable then the resulting turbulence  is able to destroy the tachocline rather fast.
\subsection{Chemical mixing}
The flow pattern of the magnetic instability also mixes passive scalars like temperature and chemical 
concentrations. The instability, therefore,  could  be relevant for the so-called lithium problem. In order to 
explain the observed  lithium concentration at the solar surface  one needs a  turbulent  mixing beneath the convection zone  
which  enhances the microscopic value of the diffusion coefficient 
of 30 cm$^2$/s by (say) two orders of magnitude. Note the smallness of this quantity; only a very 
mild  turbulence can provide such a small value of  the diffusion coefficient
\begin{equation}
D_{\rm T}\simeq \langle u'^2\rangle \tau_{\rm corr}.
\label{DT}
\end{equation}
This relation is used here as a rough estimate, a quasilinear theory of turbulent mixing has been established by  R\"udiger \& Pipin (2001) also for rotating turbulences. For a correlation time of the order of the  rotation period the desired mixing velocity is only 1 cm/s. 

\begin{figure}[h]
\begin{center} 
\includegraphics[width=8cm,height=5cm]{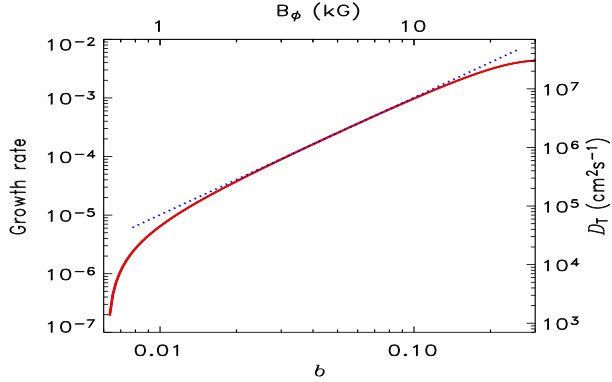}
  \caption{Growth rates in units of $\Om$ for the two-belts model and for a fixed radial scale of 1000 km. The  right-hand scale gives the estimated values for the diffusion coefficient for passive scalars and the uppermost scale gives the magnetic field amplitude in kGauss.}\label{f5}
\end{center}
\end{figure}

One can estimate the characteristic  time by $\tau_{\rm corr}\simeq l^2/D_{\rm T}$ with  $l$ as the radial scale of the instability and $D_{\rm T}\simeq 10^4$ cm$^2$/s. For the radial scale the value 1000 km has been found by Kitchatinov \& R\"udiger (2008). With this value it follows $\tau_{\rm corr} \simeq 10^{12}$ s which corresponds to a very small normalized growth rate of $ \tau_{\rm rot}/\tau\simeq 10^{-6}$. The resulting toroidal field which fulfills this condition is  smaller than 600 Gauss (Fig.~\ref{f5}). Stronger fields would produce a too strong mixing  which  would lead to much  smaller values for the lithium abundance in the solar convection zone than  observed.

Our result in connection with  the observed lithium values  also excludes the possibility that some  dynamo works in the upper part of the solar radiative core.  If such a (`Tayler-Spruit') dynamo existed then the resulting toroidal fields with less than 600 Gauss are much too weak to influence  the  magnetic activity  of the Sun with magnetic fields exceeding 10 kGauss.
\section{Tayler instability in Taylor-Couette systems}
To simplify matters we consider the pinch-type instability in a Taylor-Couette system filled with a conducting fluid which is nonstratified in axial direction. The stationary rotation law between the cylinders is  $\Om = a+b/R^2$ where $a$ and $b$ are given by the fixed rotation rates of the cylinders. In a similar way the stationary toroidal field results as   $B_\phi=AR+B/R$. In the following we have fixed the values at the cylinders to $\Om_{\rm out}=0.5 \Om_{\rm in}$ and mostly $\mu_B=B_{\rm out}/B_{\rm in}=1$ is used. The outer cylinder radius is fixed to $2R_{\rm in}$. Reynolds number Rm and Hartmann number Ha are defined as
\begin{equation}
{\rm Rm}=\frac{\Om_{\rm in}R_{\rm in}^2}{\eta}, \quad\quad\quad\quad {\rm Ha}= \frac{B_{\rm in} R_{\rm in}}{\sqrt{\mu_0 \rho \nu \eta}},
\label{rmha}
\end{equation}
the magnetic Prandtl number  here is always put to unity. 

\begin{figure}[h]
\begin{center} 
\mbox{
\includegraphics[width=6.6cm,height=5cm]{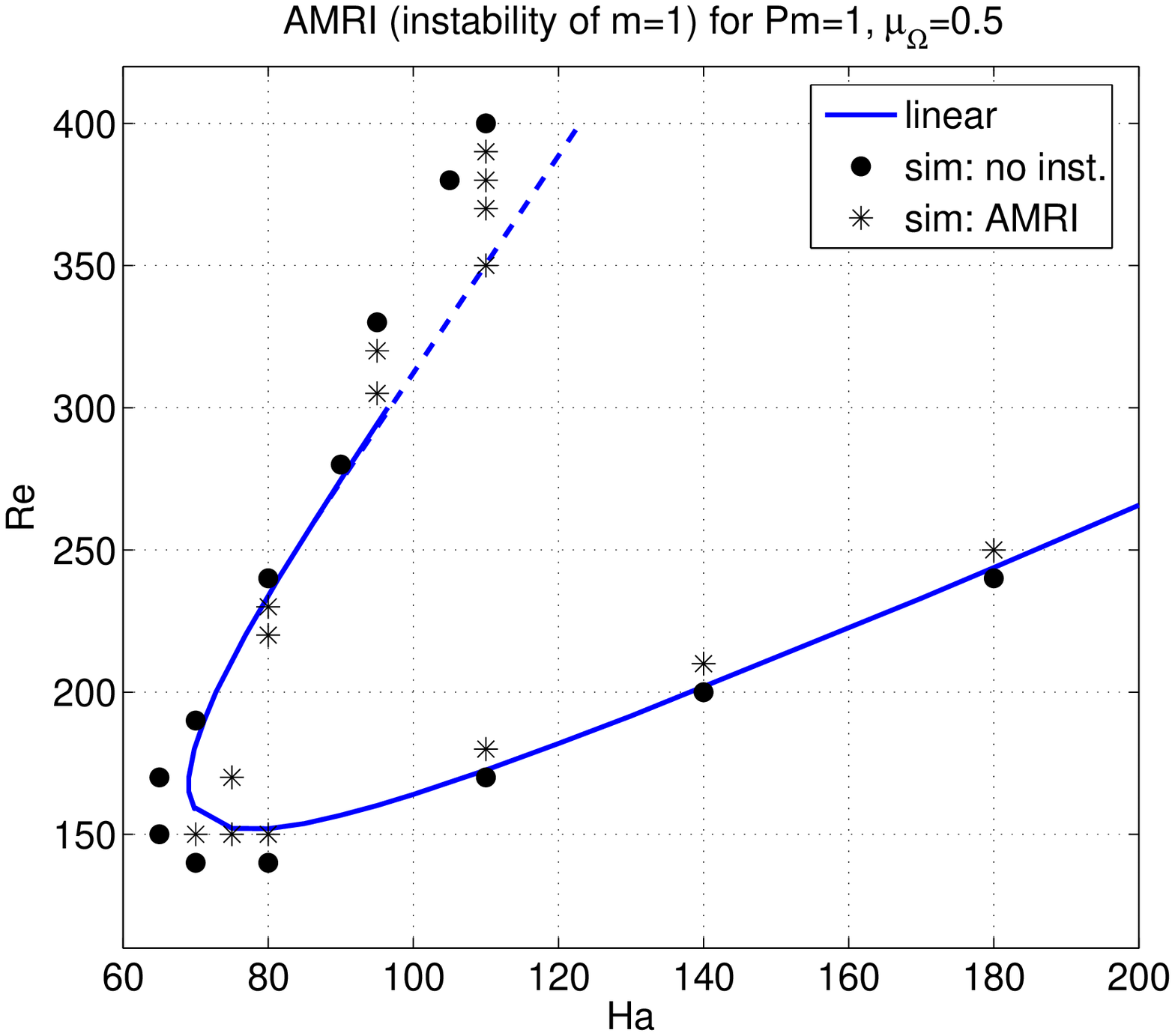}
\includegraphics[width=6.6cm,height=5cm]{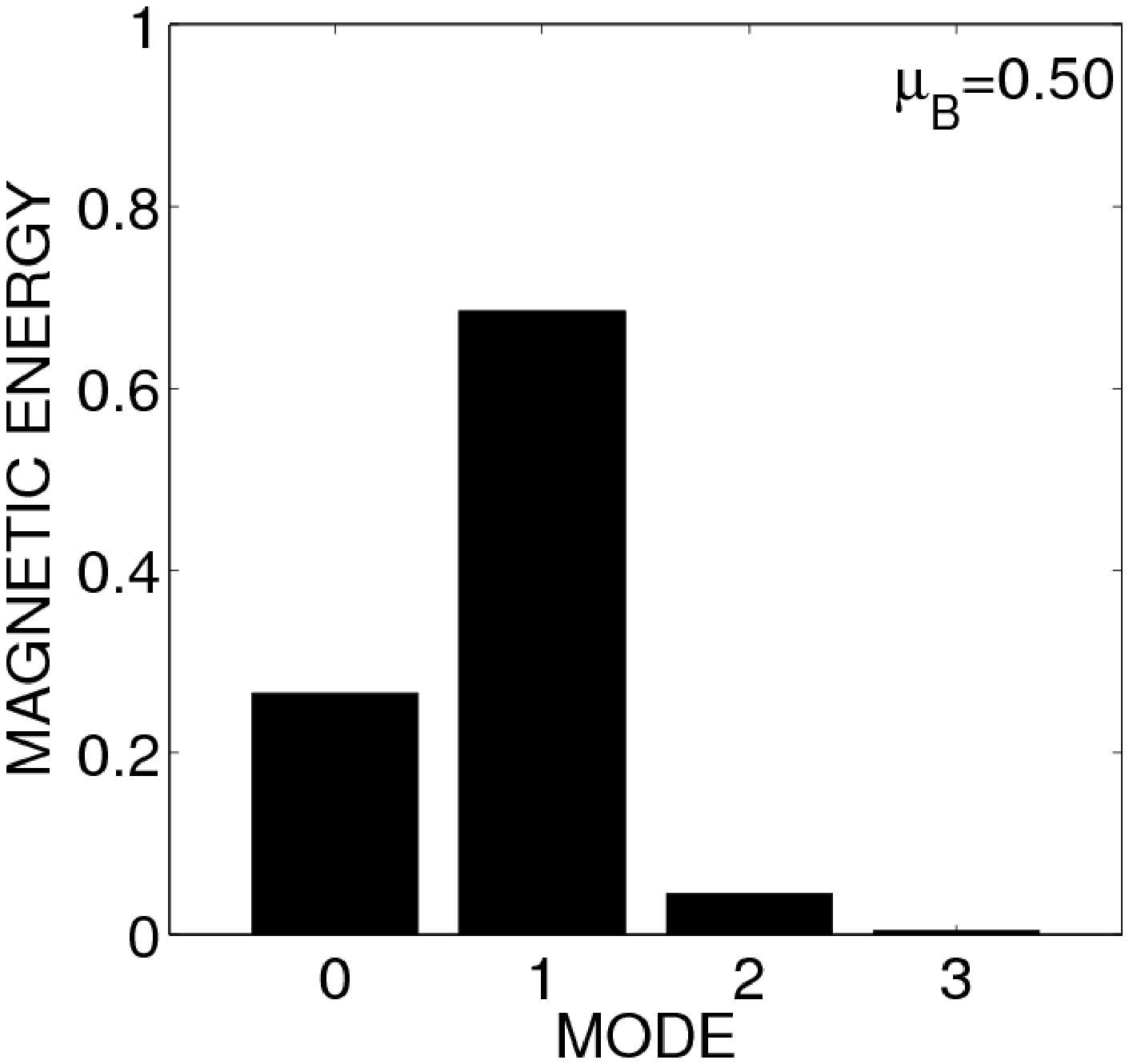}}
\caption{${\rm Pm}=1, \mu_B=0.5$ (current-free). Left: stability diagram from nonlinear simulations. Dots:  stability, stars:  instability; the solid line results from the  linear  theory.  Right: $m$-spectrum for $\rm  Re=250$, $\rm Ha=110$. The $m=1$ mode contains 69\% of the total magnetic energy. }\label{f8}
\end{center}
\end{figure}

A  detailed description of the used nonlinear MHD code for incompressible fluids is given  by Gellert et al. (2007).  In the vertical direction periodic boundary conditions are used to avoid  endplate problems. In this approximation the endplates rotate with the same rotation law as the fluid does. The height of the virtual  container is assumed as $6D$ with $D$ the gap width between the cylinders. The cylinders are considered as perfect conductors. The code was first tested for  the nonaxisymmetric AMRI which appears if stable rotation laws and stable toroidal fields (current-free, $\mu_B=0.5$) are combined (R\"udiger et al. 2007). Figure \ref{f8}
(left) shows the instability domain (solid) which for given (supercritical) magnetic field always lies between a lower Reynolds number and  an upper Reynolds number. For too slow rotation the nonaxisymmetric modes are not yet excited, but for too fast rotation the nonaxisymmetric instability modes are  destroyed so that the field becomes stable again.  

Between the limiting Reynolds numbers  the instability is no longer monochrome 
but also other  modes than   $m=1$ are nonlinearly excited. The $m=1$ mode often contains the  majority  of the total magnetic energy (Fig. \ref {f8}, right). There are also cases, however,  where $m=0$ and $m=1 $ contain nearly the same amount of magnetic energy.
Note  that the nonlinear effects can provide remarkable portions of the energy of the instability in form of axisymmetric rolls although the basic instability is a nonaxisymmetric one.
\begin{figure}[h]
\hskip-0.5cm
\mbox{
\includegraphics[width=7cm,height=5cm]{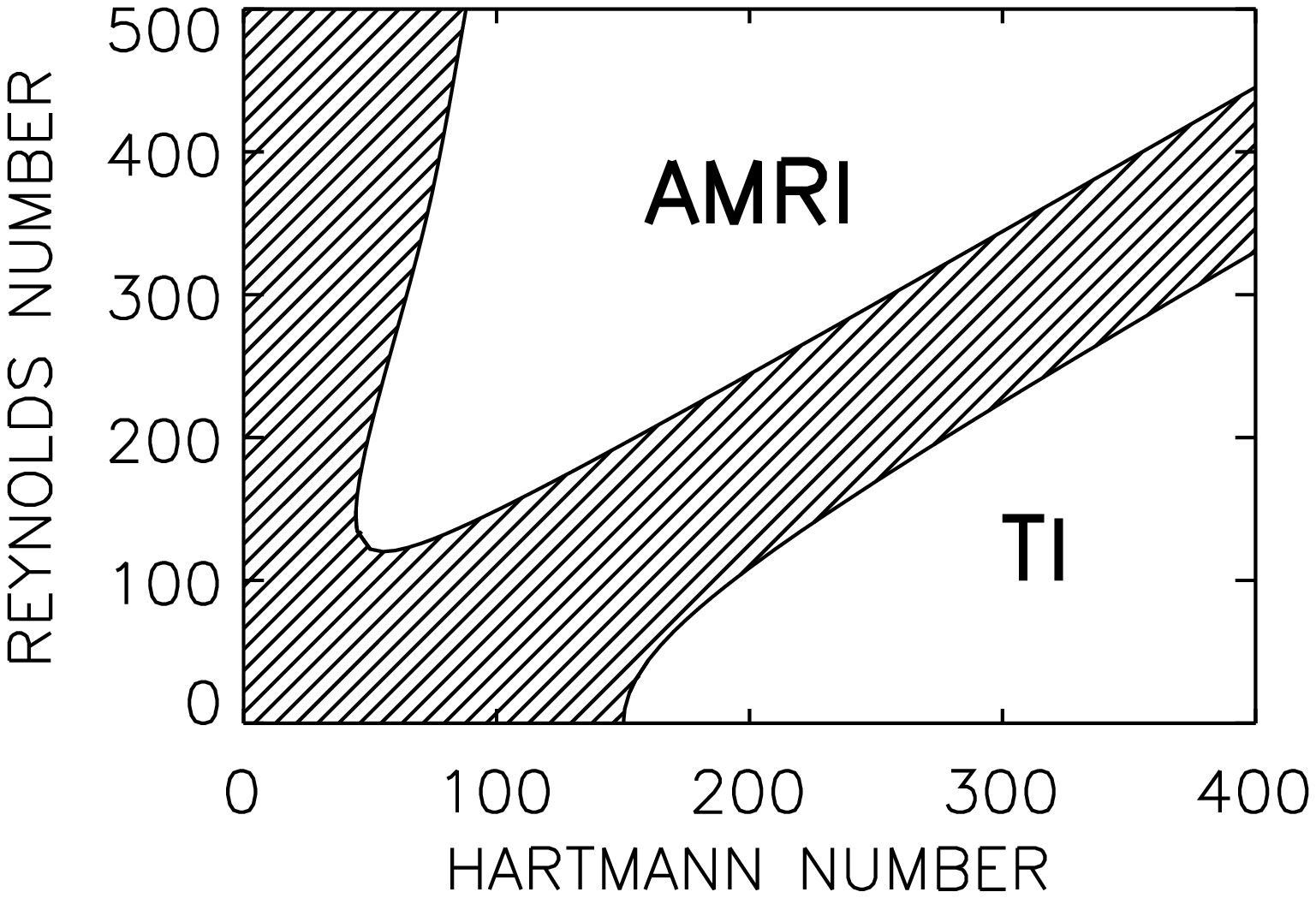}
\includegraphics[width=7cm,height=5cm]{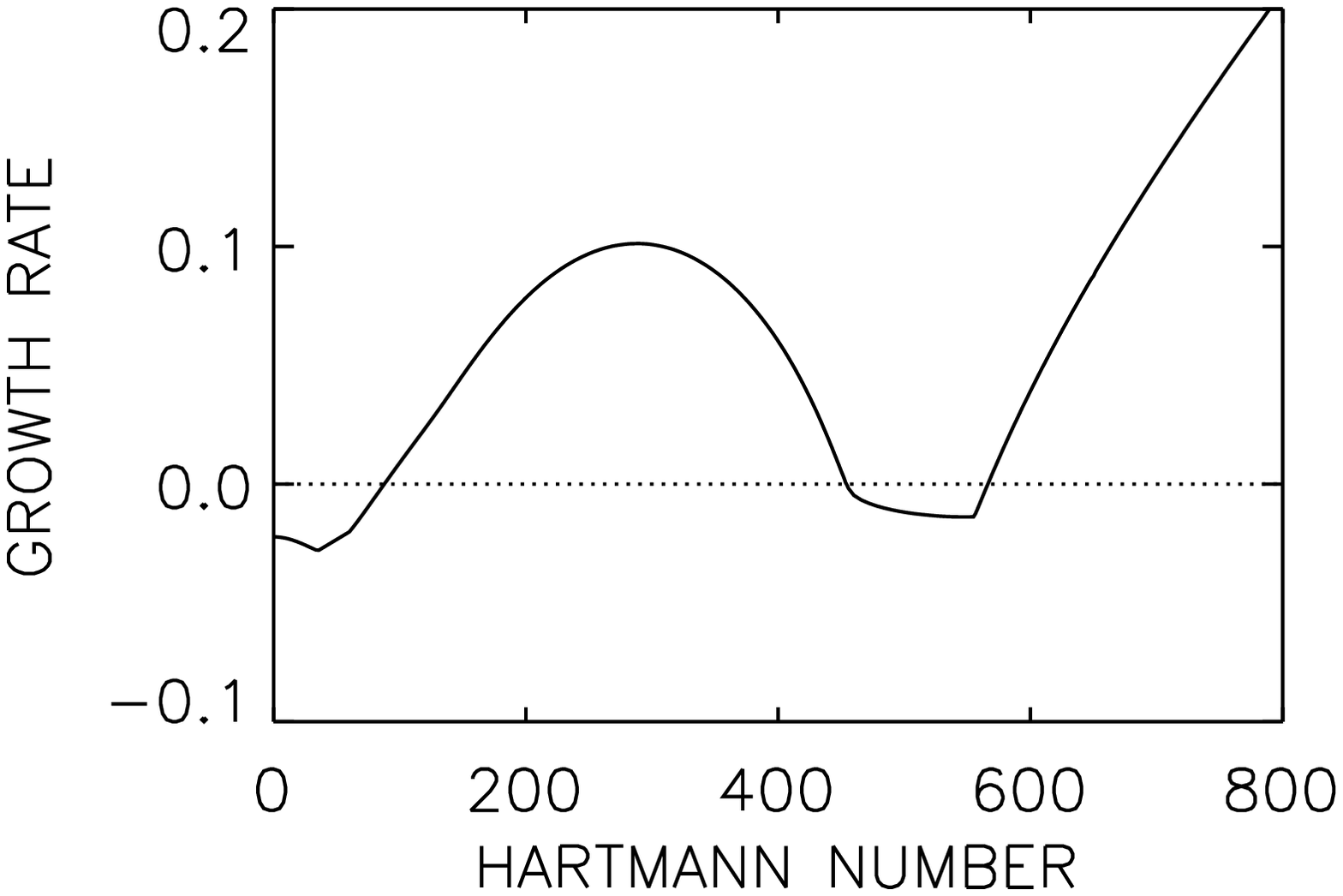}}
  \caption{Left: The instability map for  $\mu_\Omega=0.5$, $\mu_B=1$, $\rm Pm=1$. The hatched area is stable. Right: The growth rate (normalized with the rotation rate of the inner cylinder) along the horizontal line at $\rm Re=500$. The instability in the upper domain (AMRI) grows slowly while in  the lower  domain (TI) it grows much faster.}\label{f6}
\end{figure}

Now with $\mu_B=1$ positive axial currents between the cylinders are allowed. Again there are two different instability domains for the nonaxisymmetric modes with $m=1$. They are separated by a stable domain (Fig.~\ref{f6}, left). The upper one is for fast rotation and weak fields (${\rm Re}> {\rm Ha}$) and the lower one is for strong fields and slow rotation (${\rm Ha}>{\rm Re}$). The growth rates in both  domains are very different: The weak-field instability  is   slow and the strong-field instability is very  fast (Fig.~\ref{f6}, right). The rotation-dominated instability disappears for rigid rotation   while the  magnetic-dominated instability even exists  without rotation (${\rm Re}=0$). The latter one is  the Tayler instability (TI) under the stabilizing influence of the basic stellar rotation (Pitts \& Tayler 1985). The rotation-dominated (`upper') instability appears to be the AMRI which also exists under the modifying influence of weak axial currents in the fluid. Note again that i) too fast rotation finally stops both the instabilities, and ii) its growth rate  is  very small.   One must also stress that the presented results only concern the most simple case of $\rm Pm=1$. For very strong fields the stable domain between AMRI and TI disappears.

The condition for the existence of TI as given in Fig. \ref{f6} (left)  is $B_\phi > \sqrt{\mu_0 \rho} R \Om $ while the condition for  AMRI is $R \Om >B_\phi/\sqrt{\mu_0 \rho}$. Both the instabilities, however,    only exist  if the rotation is not too fast.  Nonaxisymmetric modes are {\em always  stabilized} by  sufficiently fast  rotation.

\begin{figure}[h]
\begin{center}
\mbox{ 
\includegraphics[width=6.6cm,height=4.6cm]{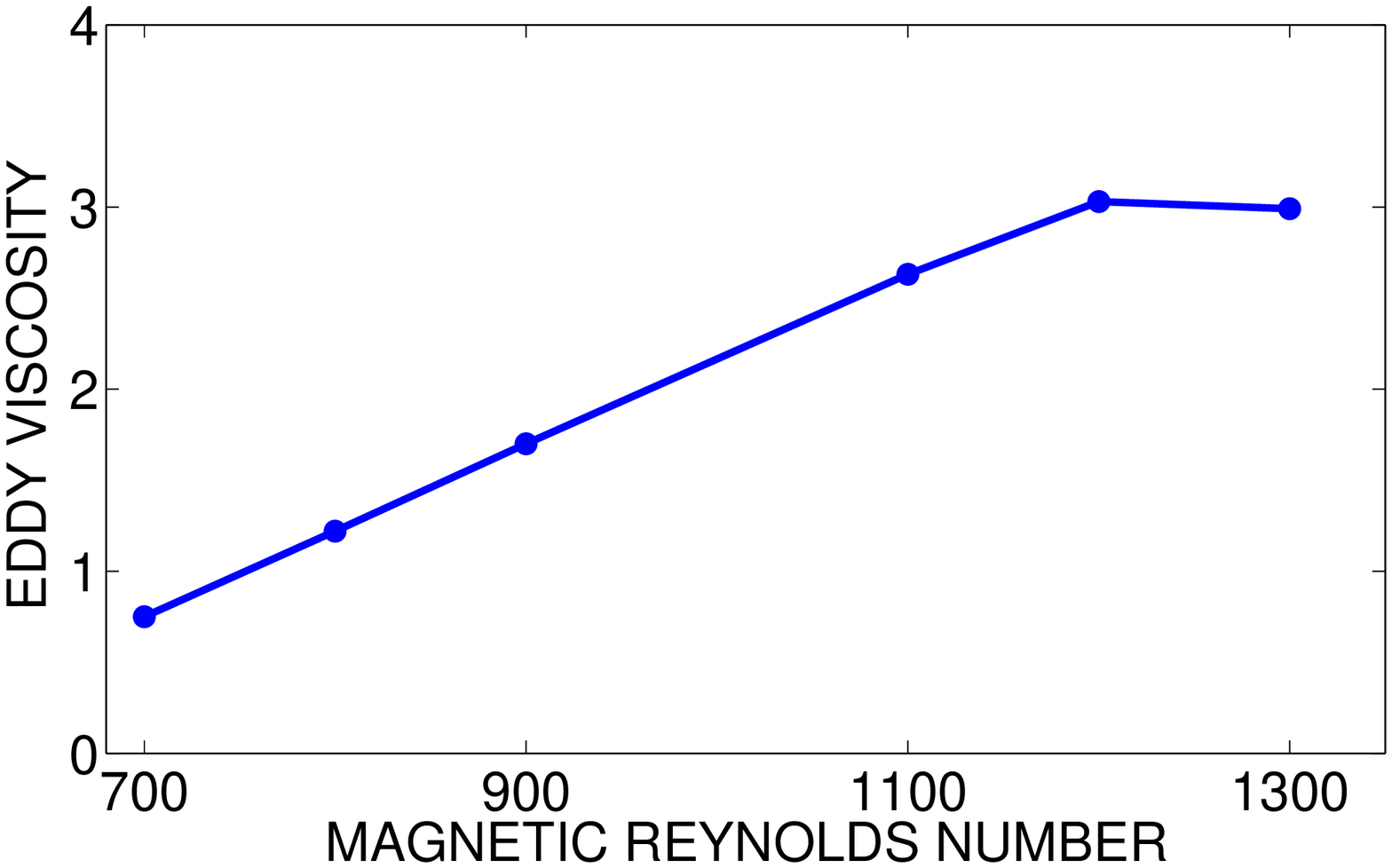}
\includegraphics[width=6.6cm,height=4.6cm]{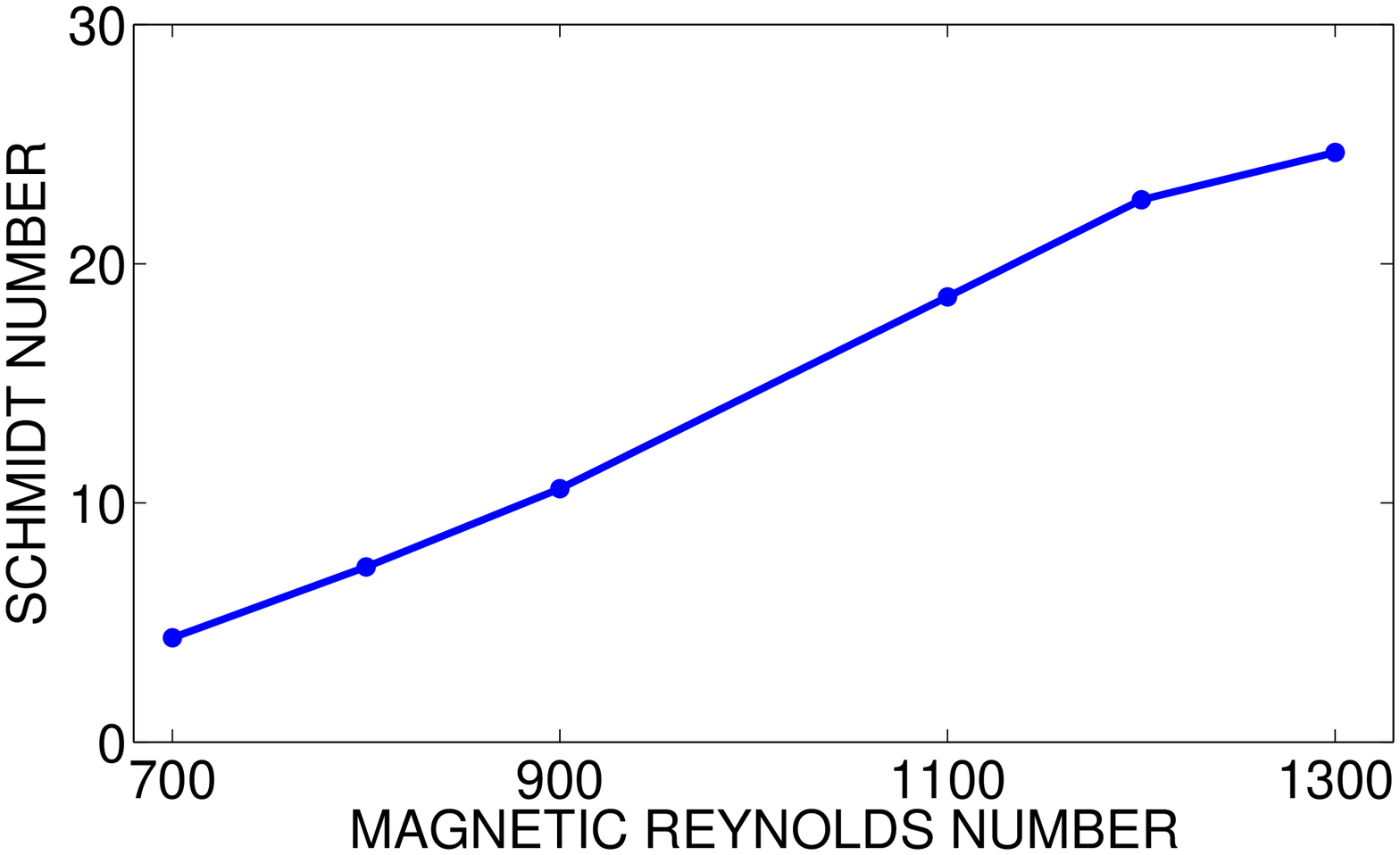}}
\caption{Left: the eddy viscosity in units of the microscopic viscosity for $\mu_\Omega=0.5$, $\mu_B=1$, $\rm Pm=1$ and $\rm Ha=500$. It grows for faster  rotation. Right: the Schmidt number (\ref{sc}).}\label{f7}
\end{center}
\end{figure}

Ap stars (with 10 kGauss and a rotation period of several  days) and neutron stars (with 10$^{12}$ Gauss and a rotation period of 10 ms) are rotation-dominated. Their instabilities are  {\rm  not} of the Tayler-type. In the following we have thus considered the AMRI in more detail. For given Ha ($=500$) the eddy viscosity,  the diffusion coefficient $D_{\rm T}$ and the Schmidt number
\begin{equation}
{\rm Sc}=\frac{\nu_{\rm T}}{D_{\rm T}}
\label{sc}
\end{equation}
are computed. The  eddy  viscosity $\nu_{\rm T}$ is the ratio of the angular momentum transport by  Reynolds stress  and Maxwell stress and the differential  rotation.    We   find  a   $\nu_{\rm T}$ of the  order  of the microscopic value (Fig.~\ref{f7}, left). The maximum exists as the instability -- as mentioned -- disappears for too fast rotation.  The averaging procedure  concerns the whole container so that the values in Fig. \ref{f7} are lower limits. 

So far the diffusion coefficient for chemicals could only be estimated by  $D_{\rm T}\simeq \langle u_R'^2\rangle/\Om$.  It proves  to be much smaller than the viscosity. Brott et al. (2008) have shown that for too strong mixing   the stellar evolution  is massively affected. A better theory   must solve the diffusion  equation. 

Accepting this approximation the resulting Schmidt number (\ref{sc}) reaches values of $20\dots30$ (Fig.~\ref{f7}, right). Obviously, the angular momentum is mainly transported by the Maxwell stress while the diffusion of passive scalars is due  to only the Reynolds stress  which is much smaller.   Quite similar  results have been  obtained by Carballido et al. (2005) and Johansen et al. (2006) for the Schmidt number of the standard MRI. Maeder \& Meynet (2005) for a hot star with 15 solar masses and for 20 kGauss find much higher values of the Schmidt number ($10^6$). Also Heger et al.  
(2005)  work with magnetic amplitudes of 10 kGauss for which $\Om_{\rm A}<\Om$ hence the growth rates are small.

{}

\vskip-0.2cm
\begin{thebibliography}{}
\bibitem[]{} Arlt, R., Sule, A. \& R\"udiger, G. 2005, \textit{A\&A} 441, 1171
\bibitem[]{} Arlt, R., Sule, A. \& R\"udiger, G. 2007, \textit{A\&A} 461, 295
\bibitem[]{} Brott, I., Hunter, I., Anders, P., Langer, N. 2008, in: B.W. O'Shea et al. (eds.), 
     \textit{FIRST STARS III},
     AIPC 990, p.\ 273 
\bibitem[]{} Cally, P.S. 2001, \textit{SoPh} 199, 231
\bibitem[]{} Cally, P.S. 2003, \textit{MNRAS} 339, 957
\bibitem[]{} Carballido, A., Stone, J.M. \& Pringle, J.E. 2005, \textit{MNRAS} 358, 1055
\bibitem[]{} Charbonneau, P., Dikpati, M. \& Gilman, P.A.  1999, \textit{ApJ} 526, 523
\bibitem[]{} Garaud, P. 2007, in: D.W. Hughes et al. (eds.), \textit{The Solar Tachocline}, p.~147
\bibitem[]{} Gellert, M., R\"udiger, G. \& Fournier, A. 2007, \textit{AN} 328, 1162
\bibitem[]{} Heger, A., Woosley, S.E. \& Spruit, H.C. 2005, \textit{ApJ} 626, 350
\bibitem[]{} Johansen, A., Klahr, H. \& Mee, A.J. 2006, \textit{MNRAS} 370, 71
\bibitem[]{} Kitchatinov, L.L. \& R\"udiger, G. 1999, \textit{A\&A} 344, 911
\bibitem[]{} Kitchatinov, L.L. \& R\"udiger, G. 2008, \textit{A\&A} 478, 1
\bibitem[]{} K\"uker, M. \& Stix, M. 2001, \textit{A\&A} 366, 668
\bibitem[]{} Maeder, A. \& Meynet, G. 2005, \textit{A\&A} 440, 1041
\bibitem[]{} Pitts, E. \& Tayler, R.J. 1985, \textit{MNRAS} 216, 139
\bibitem[]{} R\"udiger, G. \& Kitchatinov, L.L. 1997, \textit{AN} 318, 273
\bibitem[]{} R\"udiger, G. \& Pipin, V.V. 2001, \textit{A\&A} 375, 149
\bibitem[]{} R\"udiger, G. \& Kitchatinov, L.L.  2007, \textit{New J. Phys.} 9, 302
\bibitem[]{} R\"udiger, G., Hollerbach, R., Schultz, M., Elstner, D. 2007, \textit{MNRAS} 377, 1481
\bibitem[]{} Tayler, R.J. 1973, \textit{MNRAS} 165, 39
\bibitem[]{} Watson, M. 1981, \textit{GAFD} 16, 285
\end{thebibliography}
\end{document}